\documentclass{ws-ijmpe}

\begin{document}

\markboth{Andr\'as L\'aszl\'o, Zolt\'an Fodor, Gy\"orgy Vesztergombi for the NA49 Collaboration}{New results and perspectives on 
$R_{AA}$ measurements below $20\,\mathrm{GeV}$ CM-energy \dots}

\catchline{}{}{}{}{}

\title{NEW RESULTS AND PERSPECTIVES ON 
$R_{AA}$ MEASUREMENTS BELOW $20\,\mathrm{GeV}$ CM-ENERGY AT FIXED TARGET MACHINES}

\author{\footnotesize ANDR\'AS L\'ASZL\'O, ZOLT\'AN FODOR, GY\"ORGY VESZTERGOMBI}

\address{KFKI Research Institute for Particle and Nuclear Physics of the Hungarian Academy of Sciences, 
H-1525 Budapest, P.O.Box 49, Hungary\\
laszloa@rmki.kfki.hu}

\author{for the NA49 Collaboration
}

\maketitle

\begin{history}
\received{(received date)}
\revised{(revised date)}
\end{history}

\begin{abstract}
Transverse momentum spectra of $\pi^{\pm}$ at midrapidity are measured at high $p_{{}_T}$ 
in p+p and p+Pb collisions at $158\,\mathrm{GeV/nucleon}$ beam energy by the NA49 
experiment. This study is complementary to our previous results on the same 
spectra from Pb+Pb collisions. The nuclear modification factors $R_{A+A/p+p}$, $R_{p+A/p+p}$ 
and $R_{A+A/p+A}$ as a function of $p_{{}_T}$ are extracted and compared to RHIC measurements, thus providing 
insight into the energy dependence of nuclear modification. 
The modification factor $R_{A+A/p+A}$ proved to be consistent with our previous 
results on the central to peripheral modification factor $R_{CP}$.

The limitation of our current $p_{{}_T}$ range is discussed and planned future upgrades 
are outlined. Some aspects of the FAIR-CBM experiment are also presented as a 
natural future continuation of the measurements at very high $p_{{}_T}$.
\end{abstract}

\section{Introduction}

One of the most striking features observed at BNL-RHIC is the suppression of
high $p_{{}_T}$ production in central A+A collisions relative to peripheral A+A 
or p+A or p+p collisions.
This is generally interpreted as a sign of parton energy loss in hot and dense strongly interacting
matter created at the early stage of nucleus-nucleus collisions.
This interpretation implies that the suppression should decrease towards lower energies
where the initial energy and parton density is expected to be much smaller.

Numerous results on the energy dependence of hadron yields and spectra indicate that
the onset of deconfinement is located at lower SPS energies (see e.g.\ \cite{Friese,Marek}).
Existing data on central and peripheral Pb+Pb collision at $158\,\mathrm{GeV/nucleon}$ from the NA49 experiment
allow to measure the ratio $R_{CP}$ up to about $3.5\,\mathrm{GeV/c}$ in 
transverse momentum (see \cite{QM05,SQM05}). A slight suppression is seen in this range, but 
the interpretation of this result is hindered by the poorly known 
interference with the Cronin effect\footnote{An enhancement of scaled hadron yields in 
p+A collisions relative to p+p collisions is called Cronin effect. The scaling is performed 
with the number of binary collisions.} (see \cite{Cronin}). Therefore, the nuclear modification 
factors $R_{A+A/p+p}$ and $R_{A+A/p+A}$ would give a 
clearer picture: the first one is expected to contain a certain amount of 
Cronin effect and a possible suppression, while in the second quantity 
the Cronin effect approximately cancels. Therefore, 
our experiment extracted the modification factors $R_{Pb+Pb/p+p}$, $R_{p+Pb/p+p}$ and $R_{Pb+Pb/p+Pb}$ 
from the existing p+p (see also \cite{pp}), p+Pb and Pb+Pb data at top ion-SPS energy. 
The modification factor $R_{Pb+Pb/p+Pb}$ (which does not contain the Cronin effect 
to first order) confirms our previous observations with $R_{CP}$.

Our experiment, however, has only limited statistics on p+p and p+Pb collision 
data at top ion-SPS energy. Therefore, future data runs are planned. 
The proposed FAIR-CBM high luminosity experiment has also a great potential 
in this field, as it will be able to populate the momentum space up to 
the kinematic limit.

\section{Data analysis}

For the data analysis strategy of p+p and p+Pb collisions, the method 
developed earlier for Pb+Pb collisions was adopted (see \cite{QM05}). This guarantees 
a good cancellation of the already small systematic biases. The study
was performed in the rapidity interval $-0.3\leq y \leq 0.7$ 
(as in Pb+Pb) .

\begin{figure}[th]
\centerline{\psfig{file=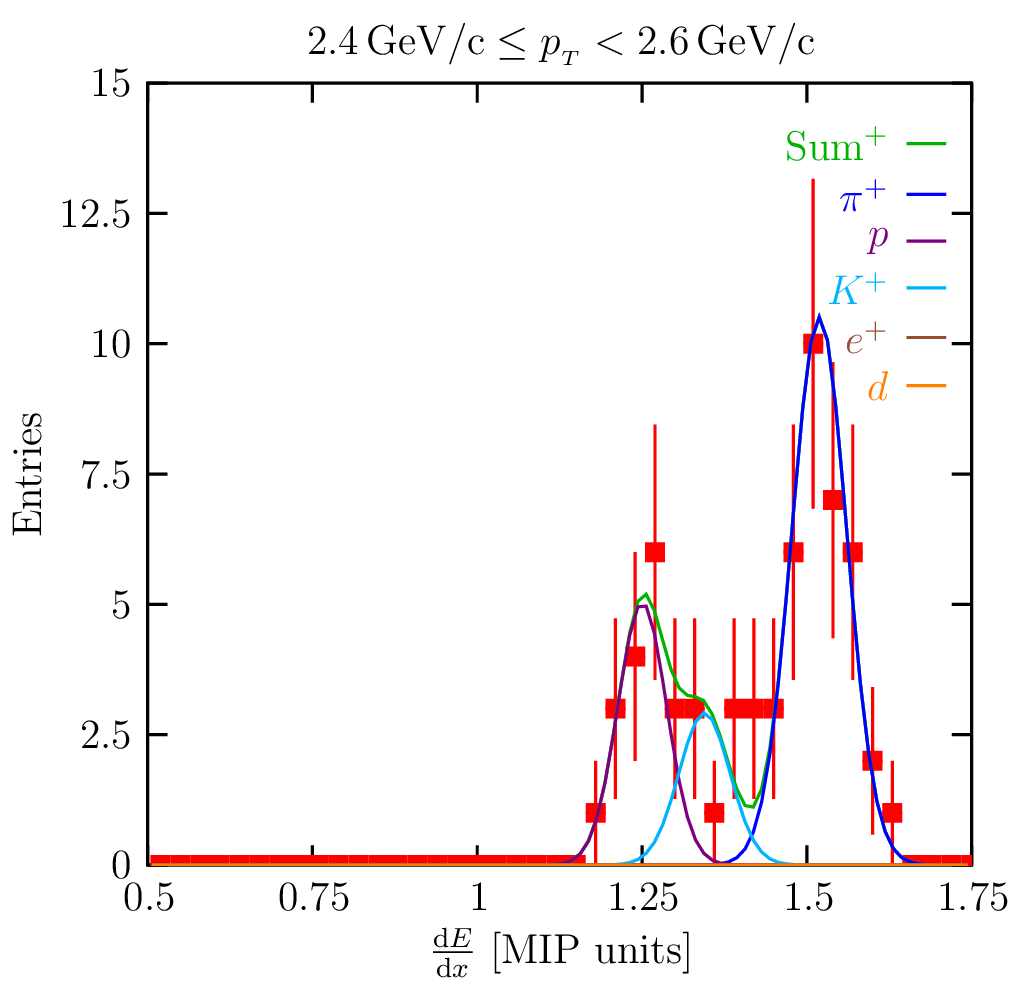,width=4.2cm}\psfig{file=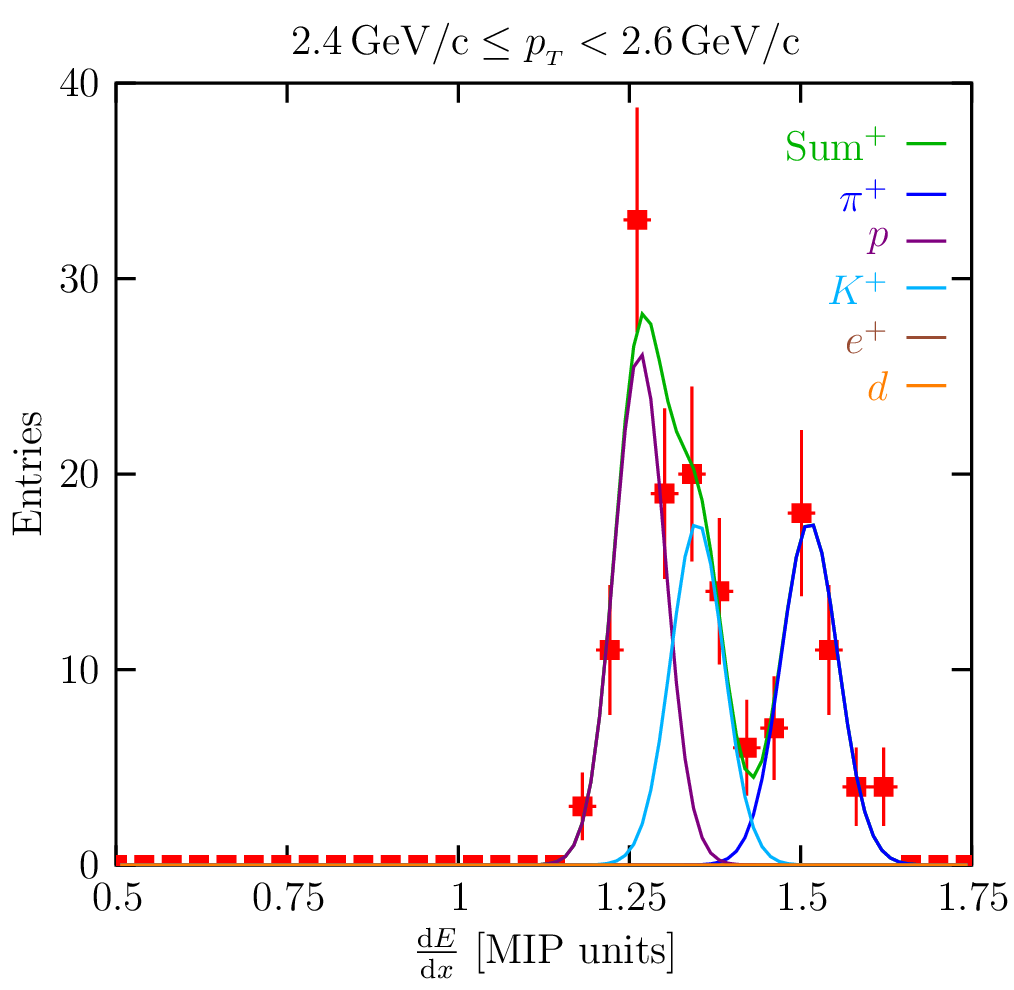,width=4.2cm}\psfig{file=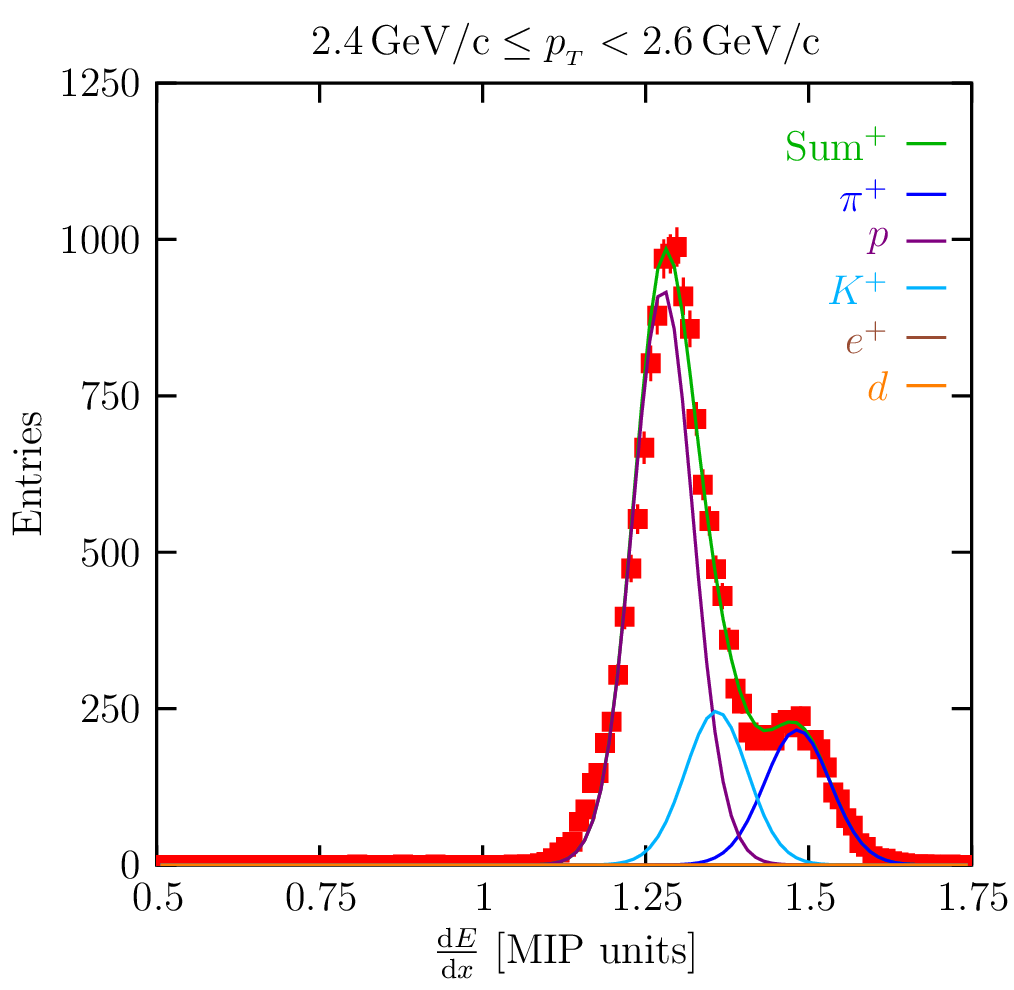,width=4.2cm}}
\vspace*{8pt}
\caption{Demonstration of the 
reliability of the inclusive particle identification by specific energy 
loss at high $p_{{}_T}$, in p+p, p+Pb and Pb+Pb(0-5\%) collisions, respectively.}
\label{dedx}
\end{figure}

The only limitation in our data analysis is our relatively poor statistics 
on p+p and p+Pb compared to Pb+Pb. This also affects our 
particle identification procedure. However, the area conservation property 
of the employed Poisson maximum-likelihood method (see \cite{BakerCousins}) for the inclusive 
disentangling of the energy loss spectra allows to make a good 
quality $\frac{\mathrm{d}E}{\mathrm{d}x}$ analysis even in the low statistics 
range of momentum space (see Fig.\ \ref{dedx}).

\begin{figure}[th]
\centerline{\psfig{file=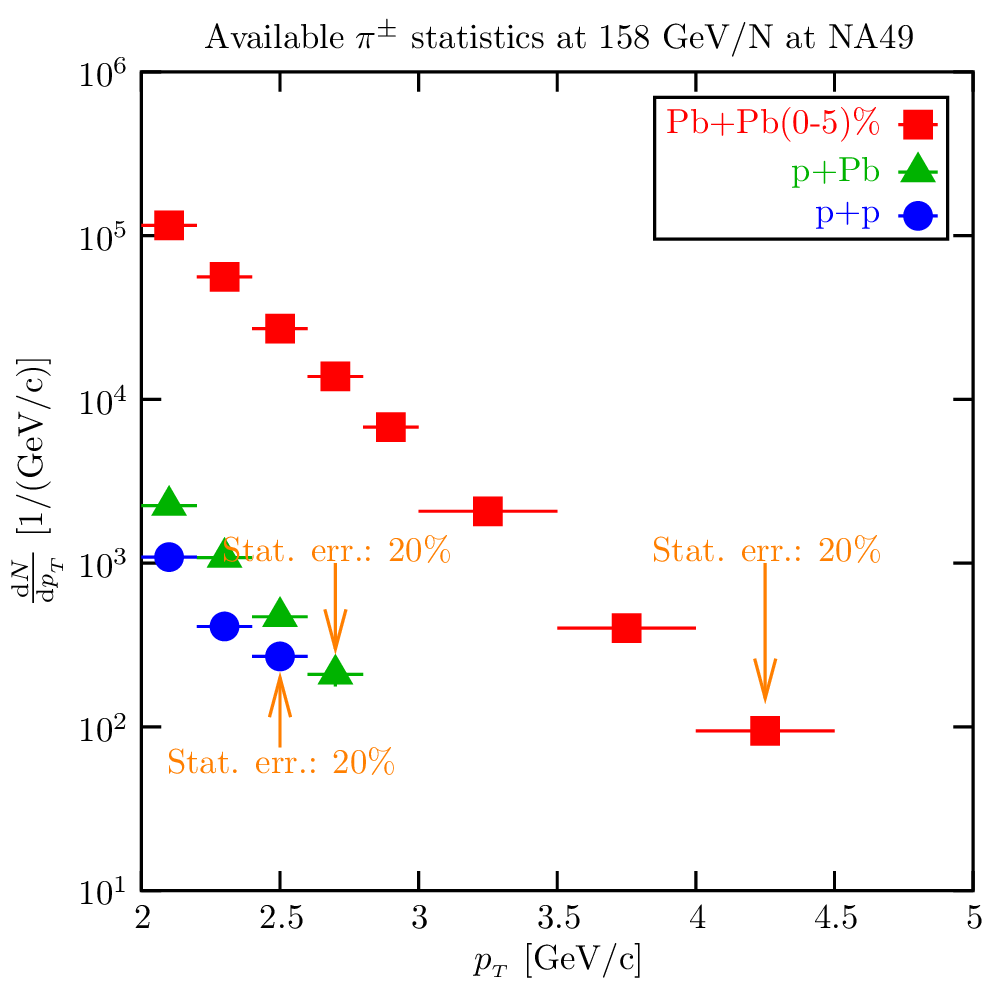,width=6cm}\psfig{file=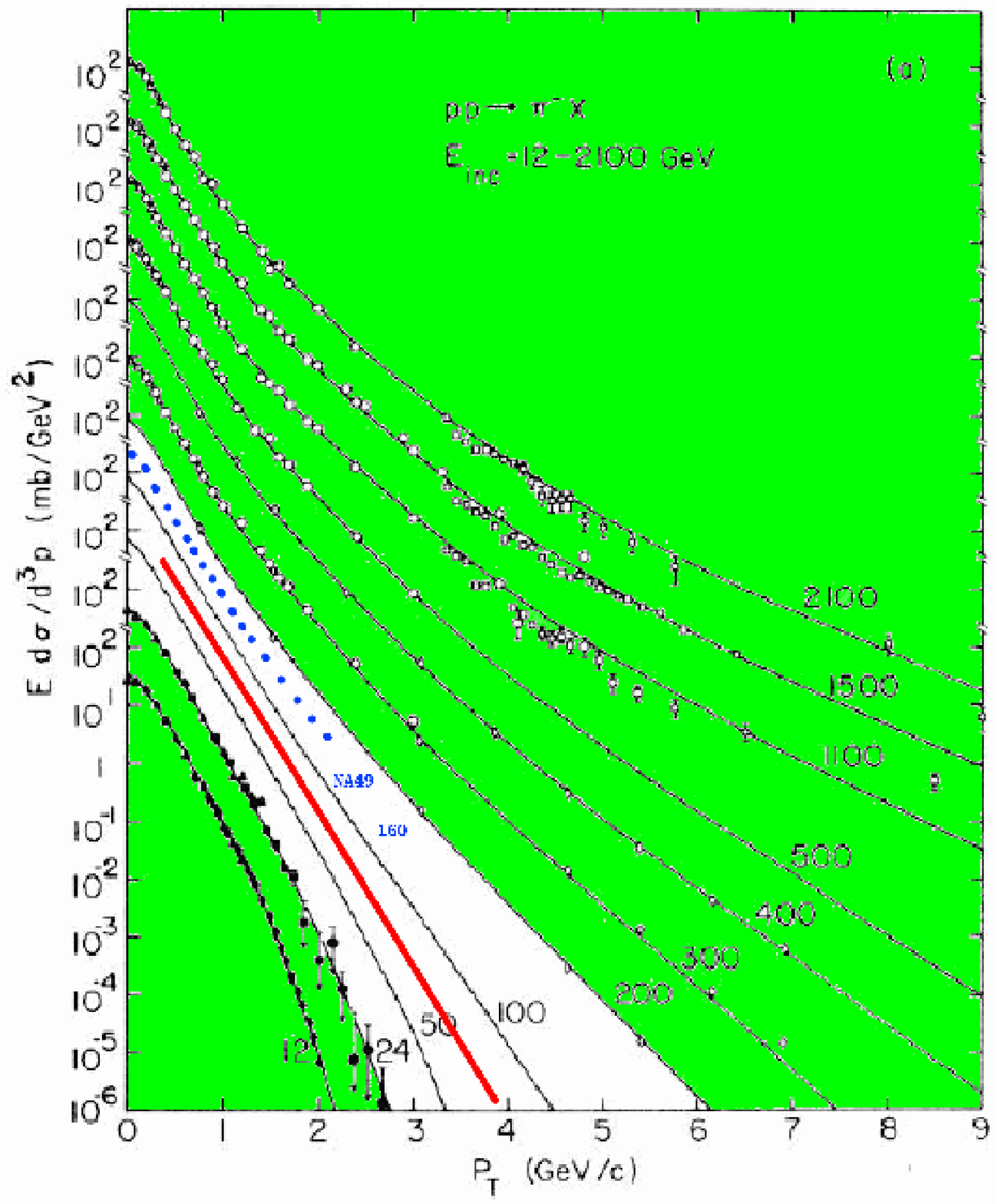,width=6cm}}
\vspace*{8pt}
\caption{Left panel: currently available 
$p_{{}_T}$ range for $\pi^{\pm}$ in p+p, p+Pb and Pb+Pb collisions at $158\,\mathrm{GeV/nucleon}$ 
at midrapidity in the experiment NA49. The p+p and p+Pb range is severely 
limited by statistics. Right panel: 
a compilation of all existing data for $\pi^{-}$ production in the p+p collisions 
in the nearby energy range, shown together with the new NA49 data. As can be 
seen there are no data near $158\,\mathrm{GeV}$, and the shape of the spectra 
changes rapidly with beam energy from convex to concave (transition shown by the solid line).}
\label{stat}
\end{figure}

For completeness, we show the currently available $\pi^{\pm}$ statistics 
in p+p, p+Pb and Pb+Pb(0-5\%) collisions at $158\,\mathrm{GeV/nucleon}$, 
and the lack of existing p+p reference data in the nearby energies 
in Fig.\ \ref{stat}. (See also \cite{Beier}.)

\section{Physics results}

\begin{figure}[th]
\centerline{\psfig{file=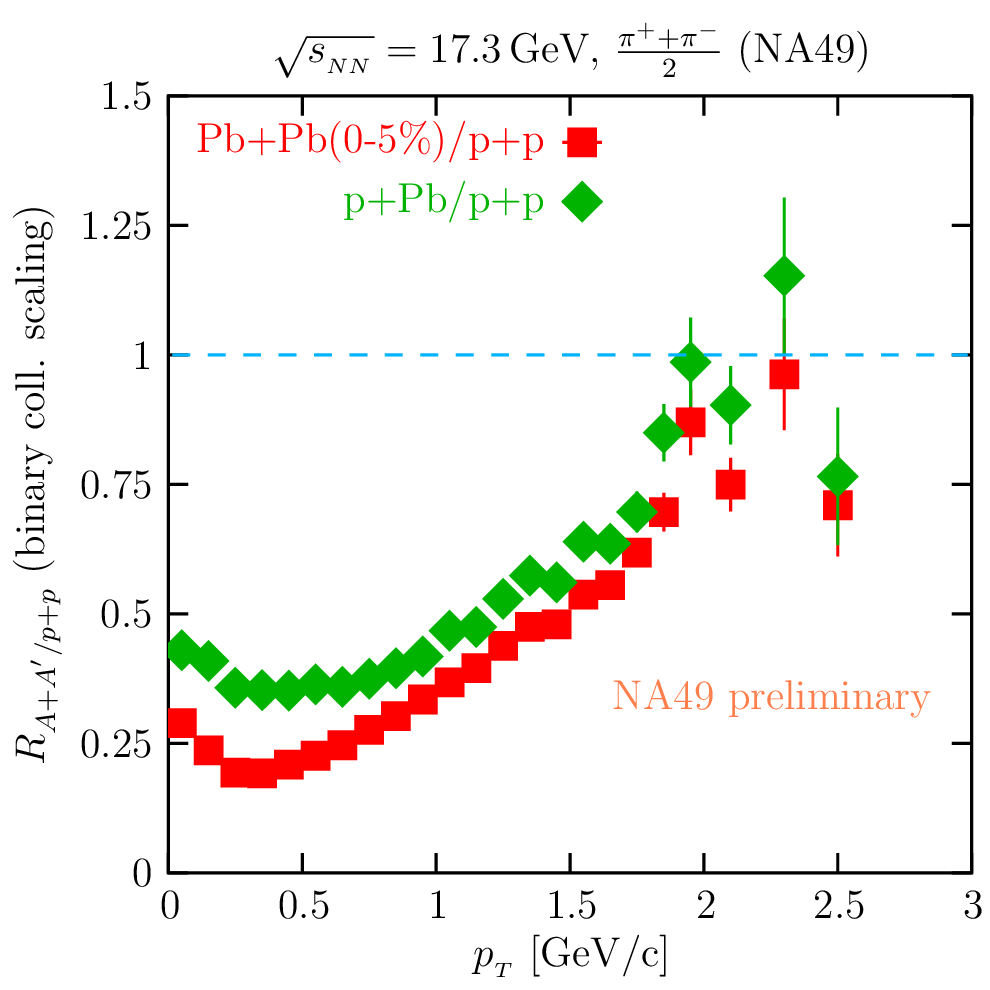,width=6cm}\psfig{file=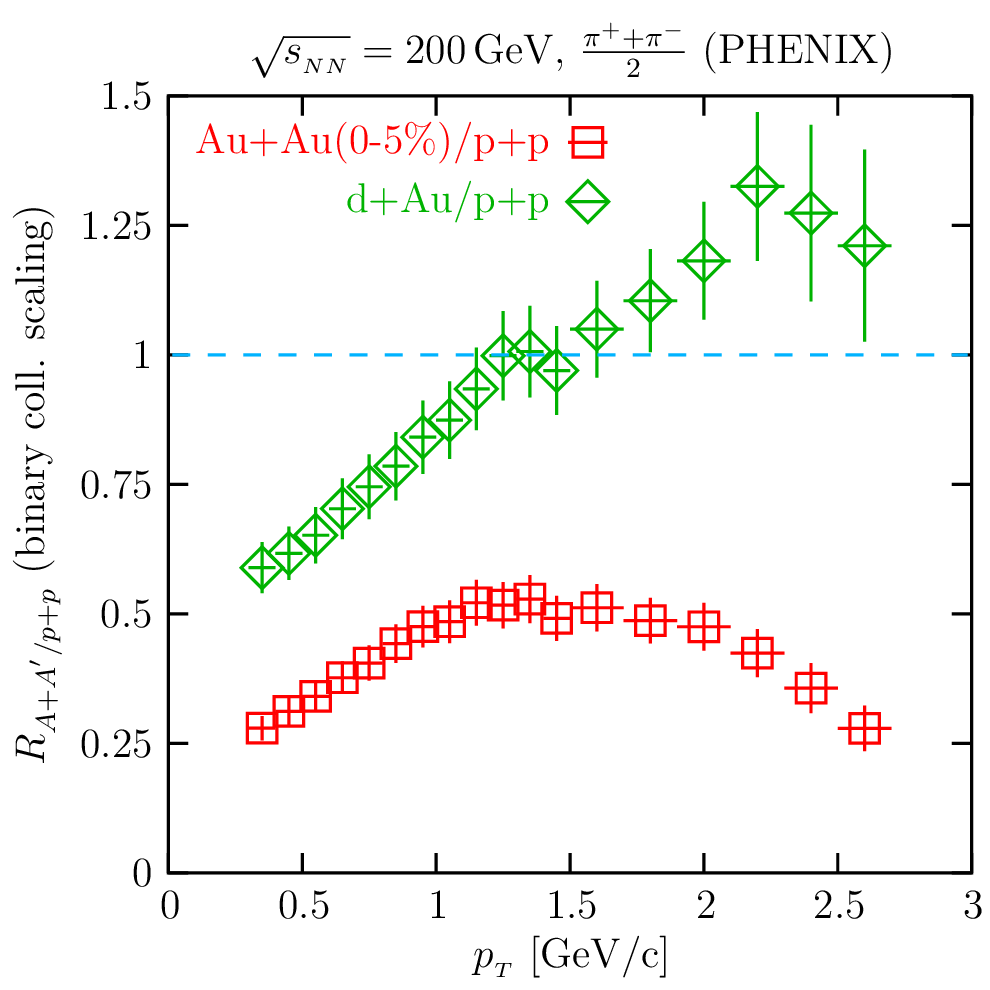,width=6cm}}
\centerline{\psfig{file=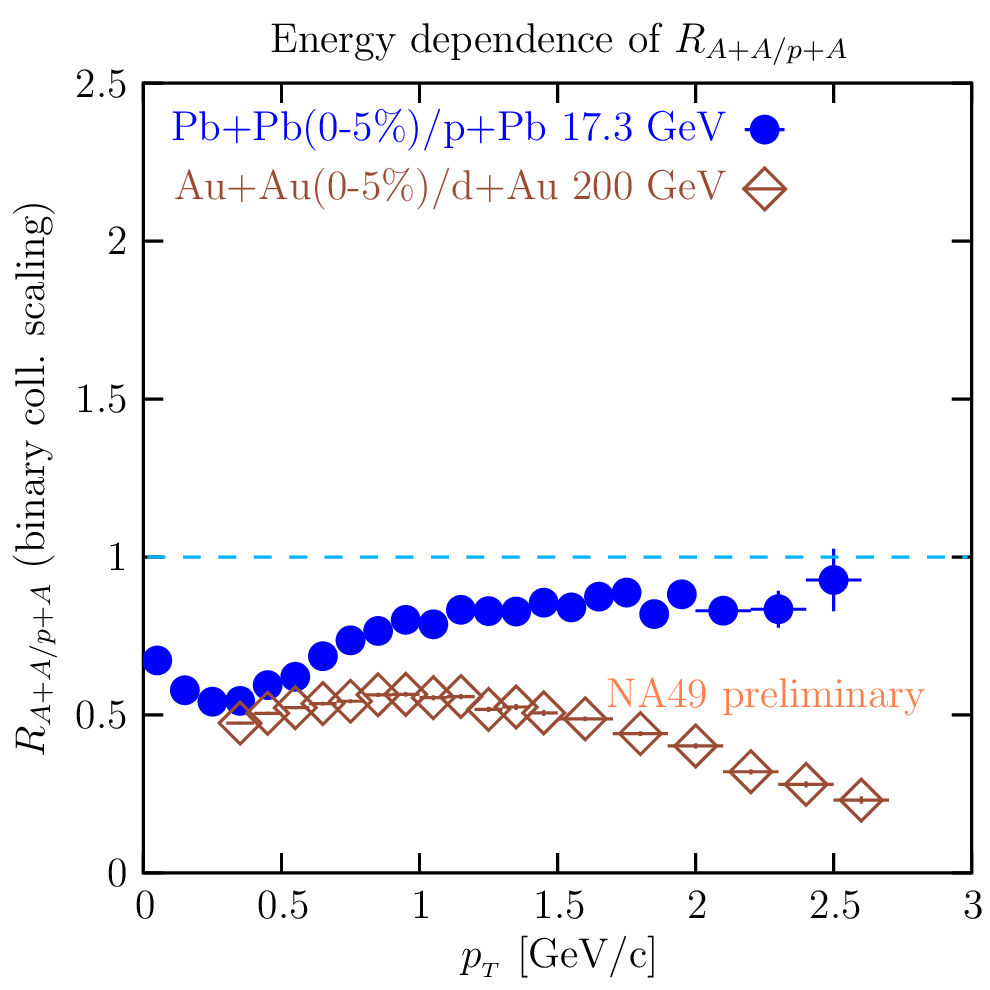,width=6cm}\psfig{file=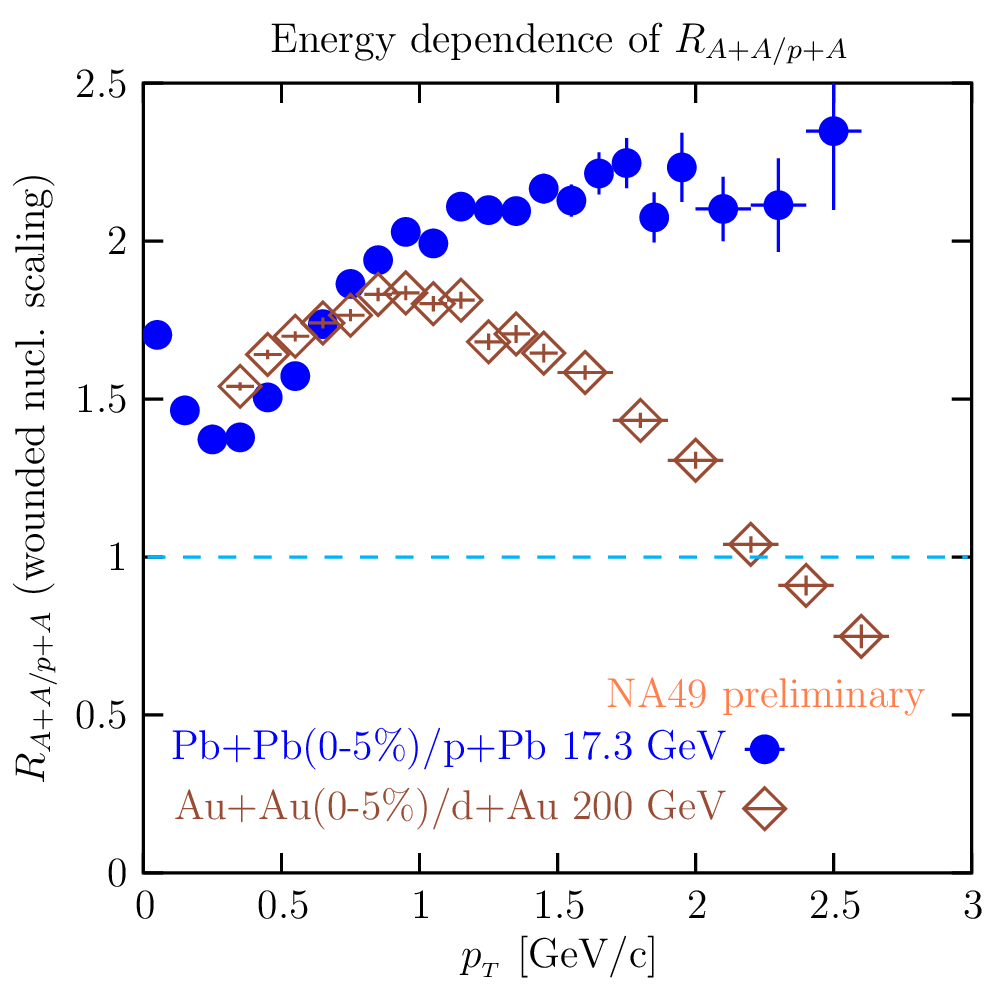,width=6cm}}
\vspace*{8pt}
\caption{Upper panels: energy dependence of the nuclear modification factor
$R_{A+A^{'}/p+p}$ for $\pi^{\pm}$ from top ion-SPS to top RHIC energies. Lower left panel: 
energy dependence of the nuclear modification factor $R_{A+A/p+A}$ for 
$\pi^{\pm}$ from top ion-SPS to top RHIC energies (binary collision scaling). 
Lower right panel: the same with wounded nucleon scaling.}
\label{edep}
\end{figure}

Fig.\ \ref{edep} shows the energy dependence of the nuclear modification factors 
$R_{A+A/p+p}$, $R_{p+A/p+p}$ and $R_{A+A/p+A}$ from top ion-SPS to top RHIC energies (see \cite{PHENIXAuAu,PHENIXdAu}). 
Strong increase of the modification factors $R_{A+A/p+p}$, $R_{p+A/p+p}$ is observed 
at top SPS energy. 
For $R_{A+A/p+A}$ (here the Cronin effect approximately cancels) 
a slight high $p_{{}_T}$ suppression is observed with binary collision scaling. 
When using wounded nucleon scaling, the $R_{A+A/p+A}$ shows an approximate 
energy-independence at low $p_{{}_T}$. Both observations confirm our 
previous similar observations on $R_{CP}$ (see \cite{QM05}). However, we are not able 
to answer the interesting question of higher $p_{{}_T}$ behavior 
(above about $2.5\,\mathrm{GeV/c}$) with the currently existing p+p and p+Pb data.

\section{Future plans}

As the current p+p and p+Pb statistics limits us to $p_{{}_T}\leq2.5\,\mathrm{GeV/c}$, 
one of our future plans is to record more p+p and p+Pb reference
data in the NA49-future experiment which is in preparation at the CERN-SPS 
to increase this range up to 4 GeV/c. Due to the limitation of this
upgraded experiment and the development possibilities of the detector there is no hope 
in the foreseeable future
to reach higher luminosity than 1000 interactions per second at CERN.
There is however an unique opportunity at the FAIR accelerator in
GSI, Darmstadt which will be ready around 2013, where the planned CBM
detector \cite{CBM} will be able to tolerate an interaction rate up to $10^{9}$ which
represents an improvement by a factor of about a million relative to the available SPS
setup.

Though the proton beam energy at the FAIR accelerator will be limited
to 90 GeV this will not decrease the physics interest, because the main issues
in p+p and p+A processes are even more exciting at high $p_{{}_T}$ at 
this energy, due to the fact 
that there is little information on either the Cronin effect or
high $p_{{}_T}$ suppression in this range. As seen from Table \ref{cbmtable}, one can 
profit from the increased luminosity despite the smallness of the estimated 
cross-sections. Even with 45 GeV beam
energy one expects significant statistics above 4 GeV/c.

\begin{table}[pt]
\begin{center}
\begin{tabular}{lllll}
\hline
\multicolumn{5}{l}{Benchmark NA49 p+p at 
$E_{\mathrm{Beam}}=158\,\mathrm{GeV}$ $30\,\mathrm{events/spill}$}\\
\hline
\hline
Events                           &  Energy [GeV]         &  $> 3\,\mathrm{GeV/c}$ & $> 4\,\mathrm{GeV/c}$ & $> 5\,\mathrm{GeV/c}$ \\
\hline
$2\cdot10^{6}$                   &  $158$                &  $100$                 & $1$                   & $0.01$                 \\
\hline
\multicolumn{5}{l}{Estimates with the assumption 
$10^{11}$ proton/sec $10^{9}$ interaction/sec}\\
\hline
$1\,\mathrm{day}=10^{14}$        &  $158$                &  $5\cdot10^{9}$        & $5\cdot10^{7}$        & $5\cdot10^{5}$         \\
\hline
\multicolumn{5}{l}{CBM Perspectives}\\
\hline
\hline
Suppression                      &  $158\rightarrow90$   &  $10^{-1}$             & $10^{-2}$             & $10^{-3}$              \\
$1\,\mathrm{day}=10^{14}$        &  $90$                 &  $5\cdot10^{8}$        & $5\cdot10^{5}$        & $500$                  \\
$20\,\mathrm{day}=2\cdot10^{15}$ &  $90$                 &  $10^{10}$             & $10^{7}$              & $10^{4}$               \\
Suppression                      &  $90\rightarrow45$    &  $10^{-3}$             & $10^{-6}$             & $10^{-10}$             \\
$20\,\mathrm{day}=2\cdot10^{15}$ &  $45$                 &  $10^{7}$              & $10$                  & $0$                    \\
\hline
\end{tabular}
\end{center}
\caption{Rate estimates and comparisons for high 
$p_{{}_T}$ at low energies.}
\label{cbmtable}
\end{table}

Referring to the "white region" in Fig.\ \ref{stat} 
between the lines of 24 and 100 
GeV beam energy, the CBM experiment will study uncharted
territory where the interesting change of the $p_{{}_T}$ spectra 
from convex to concave is happening. According to the usual argument
this phenomenon is regarded as a simple phase-space limitation effect,
but one should be aware of the fact  that at 90 GeV beam energy the 3.5 GeV/c 
transverse momentum represents only about 0.5 in $x_{{}_T}$ which 
is still rather far away from the kinematic boundary. There may be
some deeper physical effect, because the deviation from the simple
exponential spectrum already starts around 1 GeV/c. This was a big
surprise at CERN-ISR in the beginning of the seventies and due to 
lack of systematic measurements the real cause of this effect
was not elucidated since then.

Of course, this previously unimaginable luminosity which is expected
at FAIR-CBM requires very special detector, data acquisition and trigger systems.
The heart of the high $p_{{}_T}$ filtering system is the STS sub-detector.

\begin{figure}[th]
\centerline{\psfig{file=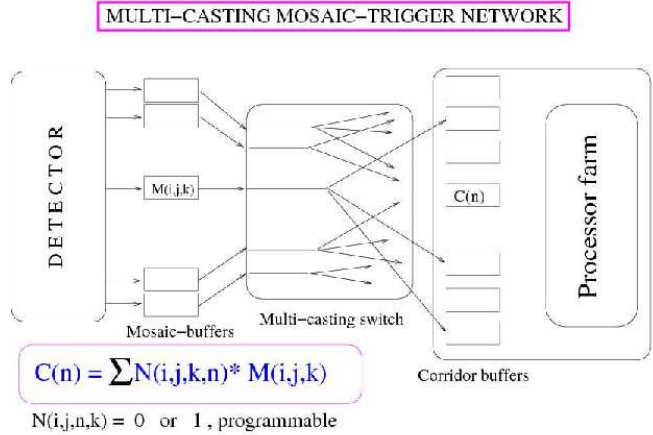,width=10cm}}
\caption{High 
$p_{{}_T}$ filtering scheme for CBM-STS subdetector.}
\label{cbmfigure}
\end{figure}

The basic idea of the filtering algorithm is that one can define rather narrow
corridors for the high $p_{{}_T}$ tracks because they are almost straight.
Due to the relatively small number of hits in these narrow 
corridors one can perform extremely fast and exhaustive searches
even in case of 1000-fold pileup. In order to ensure
full efficiency the corresponding regions of a given silicon surface (=mosaic)
can be used in a number of corridors, therefore a so-called 
"multi-tasking MOSAIC-trigger network" is proposed which is schematically
shown in Fig.\ \ref{cbmfigure}.

\section{Concluding remarks}

First NA49 results on the nuclear modification factors $R_{A+A/p+p}$, $R_{p+A/p+p}$ and 
$R_{A+A/p+A}$ were presented for the particle species $\pi^{\pm}$ 
at $158\,\mathrm{GeV/nucleon}$ beam energy, based on a study on p+p, p+Pb 
and Pb+Pb collisions.

The current statistical limitation of our p+p and p+Pb dataset is pointed out, 
and future plans are discussed. The experiment FAIR-CBM is introduced as 
a possible continuation of this low energy p+p, p+A, A+A programme at 
very high $p_{{}_T}$. Development work is in progress in the framework of the FUTURE-DAQ 
project as part of the EU FP6 contract nr. 506078 (HadronPhysics).

\end{document}